\documentclass[11pt,twoside,amsfonts,ntsha_win]{article}
\usepackage{ntsha_win}
\usepackage[OT1,T2A]{fontenc}
\usepackage[cp1251]{inputenc}
\usepackage[english,ukrainian]{babel}

\renewenvironment{abstract}{\begin{quotation}\small
\mdseries}{\end{quotation}
}
\addto\captionsukrainian{%
}

\usepackage[tbtags]{amsmath}
\usepackage{cite,amssymb,latexsym}

\providecommand\2{{^{{\boldsymbol 2}}}}
\providecommand{\abs}[1]{\lvert #1\rvert^{\mathstrut}}
\providecommand{\NU}{\lVert u\rVert^{\mathstrut\scriptscriptstyle\vphantom{1}}}
\providecommand{\us}{u_{\mathbf s}}
\providecommand{\N}[1]{\lVert u\rVert^{\mathstrut\scriptscriptstyle#1}}
\providecommand{\NT}[1]{\lVert #1\rVert^{\mathstrut\scriptscriptstyle\vphantom{1}}}
\providecommand\so{{\scriptscriptstyle 0}}
\providecommand\sv{\mathsf v}
\providecommand\bv{\boldsymbol{\mathsf v}}

\providecommand\bu{\boldsymbol{\mathsf u}}
\providecommand\buu{\boldsymbol{\dot{\mathsf u}}}
\providecommand\buuu{\boldsymbol{\ddot{\mathsf u}}}
\providecommand\uo{u_{\scriptscriptstyle 0}}
\providecommand\Wpp{{\wp}^{(1)}}
\providecommand\bwp{{\boldsymbol\wp}}
\providecommand\bwpp{{\boldsymbol\wp}^{(1)}}

\providecommand\bp{\boldsymbol{\mathsf p}}
\providecommand\Bpp{\boldsymbol{\mathsf p}^{(1)}}
\providecommand\bcd{\boldsymbol\cdot}
\providecommand\cc{\!\circ\!}
\providecommand\bx{\boldsymbol{\mathsf x}}

\title[Релятивіська дзиґа]{РЕЛЯТИВІСЬКА ДЗИҐА В ДИНАМІЦІ ОСТРОГРАДСЬКОГО}
\author[Р.~Мацюк]
{Роман МАЦЮК
}
\date{%
Інститут прикладних проблем механіки і математики\\ НАН України,\\
 вул.~Наукова, 3$^{\mbox б}$, Львів 79000
 \footnotetext{PACS~2006~numbers 11.15.Kc, 02.40.Ky, 45.20.Jj, 45.50.-j}
\footnotetext{Стаття подається в авторській редакції}}
\begin{document}


\setcounter{page}{90}  
{\renewcommand{\baselinestretch}{1.2}
 \maketitle
}
 {\hfill \small Редакція отримала статтю 20~грудня~2010~р.}

\thispagestyle{myheadings}
\markboth{~\hrulefill~Фізичний збірник НТШ т.8 2011~p.}
     {Фізичний збірник НТШ т.? 201? p.~\hrulefill~}

\begin{abstract}
Отримуємо варіяційні рівняння четвертого порядку для опису вільної релятивіської дзиґи, виходячи з рівнянь Діксона для релятивіської дипольної частки. Отриманим рівнянням надаємо однорідну просторово--часову гамільтонівську форму.
\end{abstract}

%
%
%
\chardef\atcode=\catcode`\@
\catcode`\@=11 
\def\set@low@box#1{\setbox\tw@\hbox{,}\setbox\z@\hbox{#1}%
  \setbox\z@\hbox{\dimen@\ht\z@ \advance\dimen@ -\ht\tw@
      \lower\dimen@\box\z@}%
  \ht\z@\ht\tw@ \dp\z@\dp\tw@}
\def\save@sf@q#1{{\ifhmode \edef\@SF{\spacefactor\the\spacefactor}\else
  \let\@SF\empty \fi \leavevmode #1\@SF}}
\def\@glqq{\save@sf@q{\set@low@box{''\/}\box\z@\kern-.04em\allowhyphens}}
\def\ulq{\protect\@glqq}
\def\@grqq{\save@sf@q{\kern-.07em``\kern.07em}}
\def\urq{\protect\@grqq\ }
\def\urQ{\protect\@grqq}
\catcode`\@=\atcode 

\section{ВСТУП.}
Зацікавлення таким способом опису руху буцім--класичної частки, який приводить до рівнянь з вищими похідними на
основі засобів механіки Остроградського, виникло десь біля 70-ти років тому і з тих пір не
вщухає~\cite{matsyuk:Bopp1946,matsyuk:2,matsyuk:3,matsyuk:4,matsyuk:Weyssenhoff1951,matsyuk:6,matsyuk:7}.  Останньо відновилася увага до моделей, в основі яких лежать поняття першої та
вищих кривин Френе світової ниті частки (гляди~\cite{matsyuk:8,matsyuk:9,matsyuk:10,matsyuk:11,matsyuk:12,matsyuk:LeikoMatematika}).  Побільшости, розгляд зачинається від \textit{a priori} поданого ляґранжіяну з вищими похідними, а тоді намагаються отриману динамічну систему
інтерпретувати як таку, що описує рух частки, наділеної буцім-класичним спіном (по-иньшому мовити б -- як
\textit{дзиґу}).  При цім трапляються технічні непорозуміння двох видів. По-перше, від самого початку
накладають декотрі неголономні в'язі. Ці в'язі вибираються таким чином, щоб наперед забезпечити умову, згідно
з якою ляґранжіян записується в системі координат рухомого репера~\cite{matsyuk:13}. Але ж, як вказано в
праці~\cite{matsyuk:14}, неголономні в'язі вимагають делікатнішого підходу.  Зокрема, зв'язана система втрачає
властивість варіяційности.  По-друге, відоме і звабливе припущення про унітарність вектора чотиривимірної
швидкості часом вносять запізно -- уже після того, як варіяційна процедура з певним \textit{незв'язаним}, але
й \textit{параметрично-неінваріянтним} варіяційним завданням вже зостала переведена (пор.~\cite{matsyuk:Riewe}). Такий підхід
зазнавав справедливої критики з боку ріжних авторів (гляди~\cite[стор.~149]{matsyuk:Rund}, або~\cite{matsyuk:17}). З иньшого боку, для опису релятивіської дзиґи
служать давно встановлені рівняння третього порядку Матісона~\cite{matsyuk:Mathisson1937}, рівняння другого порядку Матісона--Папапетру~\cite{matsyuk:18}, і система рівняннь першого
порядку Діксона~\cite{matsyuk:Dixon}. І ось, у
1945\textsuperscript{му} році, у зв'язку з працею Матісона~\cite{matsyuk:Mathisson1937}, Вайсенхоф з Рабе ствердили  таку думку: \ulqРівняння руху матеріяльної частки, наділеної спіном,
не збігаються з нютонівськими законами руху навіть для вільної частки в ґалілеєвській області; остається додатковий член, залежний від нутрішнього моменту, або ж спіну частки, який \textit{підвищує порядок цього диференційного рівняння до третього}.\urQ\footnote{``The equations of motion of a material particle endowed with spin do not coincide with the Newtonian laws of motion even for a  free particle in Galileian domains; there remains an additional term depending on the internal angular momentum or spin of the particle which raised the order of these differential equations to three.'' (Праця~\cite{matsyuk:2} доповідалася на засіданні Краківського відділення Польського фізичного товариства 28~лютня~1945~року.)} До наведеної думки
можемо додати, що процедура \textit{повного} усунення спінових змінних підвищує порядок диференційного
рівняння для світової ниті частки \textit{до цифри чотири.} В наступному відділі покажемо, як оце диференційне
рівняння четвертого порядку випливає з процедури усунення змінних спіну із системи рівнянь Діксона в
пласк\'ому просторі і запропонуємо вираз для функції Ляґранжа, при якій параметрично-інваріянтне варіяційне
завдання видасть світові лінії частки з нутрішнім моментом без будь-яких в'язей, що накладалися-б
\textit{перед} переведенням варіяційної процедури. В'язь \textit{постійности кривини Френе} повинна накладатися
\textit{після} переведення варіяції, і саме тому називаємо запропоновану нами функцію Ляґранжа
\textit{покриваючим ляґранжіяном.} Опісля збудуємо систему узагальнених гамільтонівських рівнянь, що
відповідатимуть цьому ляґранжіянові.

\section{ПІДВИЩЕННЯ ПОРЯДКУ І СКОРОЧЕННЯ КІЛЬКОСТІ ЗМІННИХ.}

\subsection{Рівняння Матісона--Папапетру--Діксона з умовою Матісона--Пірані: перше підвищення порядку.}
{\let\bf\relax\let\rm\relax
Аби почати з найнижчого диференційного порядку рівнянь, згадаймо рівняння Діксона для буцім-класичної частки
зі спіном, взагалі кажучи, у ґравітаційному полі,
\begin{equation}\label{matsyuk:Dixon}
\left\{
 \begin{aligned}
        \frac{D\mathcal P_\alpha}{d\tau} & =  -\frac12\,R_{\alpha\beta}{}^{\rho\nu}\dot x^\beta
    S_{\rho\nu} \\[3.5\jot]
\frac{DS_{\alpha\beta}}{d\tau} & =  \phantom{-}\mathcal P_\alpha\dot x_\beta-\mathcal P_\beta\dot x_\alpha\,,
    \end{aligned}
\right.
\end{equation}
записані з допомогою поняття коваріянтного упохіднення $\frac{D}{d\tau}$ вздовж
світової ниті з довільним в\'ідміром міркою $\tau$. В загальній теорії відносности ці рівняння мають виконуватися уздовж світової
ниті буцім-класичної частки, наділеної нутрішнім моментом кількости руху (т.~зв. ~\ulqспіном\urQ)
$S_{\alpha\beta}+S_{\beta\alpha}=0$, відповідальним за її дипольну структуру.

З-поміж кількох додаткових умов, які додаються до системи рівнянь~(\ref{matsyuk:Dixon}) аби усунути її недоозначеність (гляди~\cite{matsyuk:B/R}), ми зупинимось на умові, обраній Матісоном~\cite{matsyuk:Mathisson1937}
\begin{equation}\label{Matsyuk:4}
\dot x^\rho S_{\rho\alpha}=0.
\end{equation}
За цієї умови величина $m=\frac{\mathcal P\cdot u}{\NU}$, де $u=\dot x$, є інтеґралом руху рівно ж як і
величина скаляру нутрішнього моменту
$\sigma^{\mathbf2}=\sigma_\alpha\sigma^\alpha=S_{\alpha\beta}S^{\alpha\beta}$, де
\begin{equation}\label{matsyuk:sigma}
\sigma_\alpha=\frac{\sqrt{\abs g}}{2\NU}\varepsilon_{\alpha\beta\rho\nu}u^\beta S^{\rho\nu}\,.
\end{equation}

Умова Матісона дозволяє розв'язати співвідношення~(\ref{matsyuk:sigma}) щодо тензора спіну:
\begin{equation}\label{Matsyuk:18}
S_{\alpha\beta}=\frac{\sqrt{\abs g}}{\NU}\varepsilon_{\alpha\beta\rho\nu}u^\rho\sigma^\nu\,,
\end{equation}
і тепер сама вона набирає вигляду
\begin{equation}\label{matsyuk:5}
\sigma\cdot u=\sigma_\alpha u^\alpha=0\,.
\end{equation}

Попереднім дослідженням~\cite{matsyuk:NTSh2006} встановлено, що система рівнянь(\ref{matsyuk:Dixon}), обмежена умовою(\ref{Matsyuk:4}), є рівнозначною з такою системою:
\begin{gather}\label{Matsyuk:19}
\begin{split}
\varepsilon _{\alpha\beta\rho\nu} \ddot {u}^{\beta}u^{\rho}  {{\sigma}} ^{\nu} -
3
\,
{\frac{{{\rm {\bf \dot {u}}} \cdot {\rm {\bf u}}}}{{{\rm {\bf u}}\2}}}
\,
\varepsilon _{\alpha\beta\rho\nu} \dot {u}^{\beta}u^{\rho}  {{\sigma}} ^{\nu} +
{\frac{{m}}{{\sqrt {{\abs g}} }}}{\left[ {\left( {{\rm {\bf \dot {u}}}
\cdot {\rm {\bf u}}} \right)u_{\alpha} - {\rm {\bf u}}\2\dot {u}_{\alpha}}  \right]}
 \\
= {\frac{{{\rm {\bf u}}\2}}{{2}}}
R_{\alpha\beta}{}^{\kappa\mu} \varepsilon _{\kappa\mu\rho\nu} u^{\beta}u^{\rho}  {{\sigma}}^{\nu}
\end{split}
 \\
{\rm {\bf u}}\2{\rm {\bf \dot {  {{\sigma}}} }} + \left( {{\rm {\bf   {{\sigma}}} } \cdot {\rm
{\bf \dot {u}}}} \right)\;{\rm {\bf u}} = 0
\label{Matsyuk:20}
\\
{\rm {\bf   {{\sigma}}} } \cdot {\rm {\bf u}} = 0\,,
\label{Matsyuk:21}
\end{gather}
де \ulqпоступальна частина\urq (\ref{Matsyuk:19}) тепер вже містить третю похідну від координати частки.

Система рівнянь (\ref{Matsyuk:19}, \ref{Matsyuk:20}) є нев\'ідмірною (в\'ідмірно--байдужою), сиріч інваріянтною щодо будь-яких перетворень незалежної змінної~$\tau$, яка служить міркою уздовж світової ниті частки.

Відновлення системи рівнянь(\ref{matsyuk:Dixon}, \ref{Matsyuk:4}) досягається впровадженням змінної~$\mathcal P$ взором
\begin{equation}\label{P}
\mathcal P_\alpha=\frac{m}{\NU}\,u_\alpha+\frac{\sqrt{\abs g}}{\N3}\varepsilon_{\beta\rho\nu\alpha}\dot u^\beta u^\rho\sigma^\nu\,,
\end{equation}
якого можна записати, використовуючи позначку двоїстого тензора, так:
\begin{equation}\label{P*}
\mathcal P=\frac{m}{\NU}\,u+\frac1{\N3}\,*\,\dot u\wedge u\wedge\sigma\,.
\end{equation}

Квадрат скалярної величини вектора кількости руху~$\mathcal P$ гарно виражається через поняття першої кривини Френе світової ниті частки,
\begin{equation}\label{matsyuk:k in u}
k=\frac{\NT{\dot u\wedge u}}{\N3}\,,
\end{equation}
ось яким чином:
\begin{equation}\label{P*2}
\begin{split}
\mathcal P\2 & \overset{\mathrm{def}}= \mathcal P\cdot\mathcal P  = m^2+\frac{1}{\N6}(\,*\,\dot u\wedge u\wedge\sigma)\2  = (\,\dot u\wedge u\wedge\sigma\;\cdot\;\dot u\wedge u\wedge\sigma\,)\\
&  =  m^2+\frac{\sigma\2}{\N6}\left[(\dot u\cdot u)^2    -\dot u\2u\2\right]
  \\
 &\mspace{95mu}+\frac{(\sigma\cdot \dot u)^2\,u\2}{\N6}
    +\frac{(\sigma\cdot  u)^2\,\dot u\2}{\N6}-2\,\frac{(\dot u\cdot u)(\sigma\cdot  u)(\sigma\cdot  \dot u)}{\N6}
 \\
 & = m^2-\sigma\2k^2+\frac{1}{\N6}\left[(\sigma\cdot\dot u)\,u-(\sigma\cdot u)\,\dot u\right]\2\,.
\end{split}
\end{equation}
}
\subsection{Рівняння четвертого порядку для вільної релятивіської дзиґи.}

Надалі кладемо $R_{\alpha\beta}{}^{\mu\nu}=0$. В пласк\'ому просторі чотири-вектор спіну $\sigma$ є постійним.
Це можна угледіти, наприклад, стявши рівняння~(\ref{Matsyuk:19}) з вектором $\sigma_\alpha$ і опісля зиркнувши на рівняння~(\ref{Matsyuk:20}).

Пробуватимемо далі усувати величини $\sigma^\nu$ зі системи рівнянь (\ref{Matsyuk:19}, \ref{matsyuk:5}). Щоби
спростити підрахунки, варто обрати міркою уздовж світової ниті частки, як звичайно, натуральну мірку
$s$ так, що $\dot x_{\mathbf s}\cdot\dot x_{\mathbf s}=1$. Негайно отримуємо, використовуючи поняття двоїстого
тензора, ось яку форму рівняння~(\ref{Matsyuk:19})
\begin{equation}\label{matsyuk:7}
*\,(\ddot\us\wedge\us\wedge\sigma)+m\,\dot\us=0\,.
\end{equation}
Це рівняння має перший інтеґрал -- квадрат першої кривини Френе світової ниті
\begin{equation}\label{matsyuk:k2=dotu2}
k^2=\us\cdot\us\,.
\end{equation}
Виходячи зі взору~(\ref{P*2}), негайно бачимо, що і квадрат скалярної величини вектора кількости руху~$\mathcal P\2$ є сталим на розв'язках системи рівнянь (\ref{matsyuk:7}, \ref{Matsyuk:21}) в нашому пласк\'ому просторі.

Тепер згорнімо векторне рівняння~(\ref{matsyuk:7}) з тензором $*(\,\us\wedge\,\sigma)$, пам'ятаючи про
умову~(\ref{Matsyuk:21}). Після певних алґебричних  маніпуляцій отримаємо
\begin{equation*}
\sigma^{\mathbf2}(\ddot\us+k^2\us)=-\,m\,*\,(\dot\us\wedge\us\wedge\sigma)\,.
\end{equation*}
Оце упохіднюючи і опісля підставляючи праву частину з рівняння~(\ref{matsyuk:7}), остаточно отримуємо рівняння
\begin{equation}\label{matsyuk:8}
\dddot\us+\left(k^2-\frac{m^2}{\sigma^{\mathbf2}}\right)\,\dot\us=0\,.
\end{equation}

Коли ж тепер запровадимо позначку $\omega^2=-\frac{\mathcal P^{\mathbf2}}{\sigma^{\mathstrut\mathbf2}}$, де~$\omega$ несе
фізичне навантаження поняттям частоти осциляцій, отримаємо бажане рівняння четвертого порядку для світової
ниті вільної релятивіської дзиґи:
\begin{equation}\label{matsyuk:9}
\dddot\us+\omega^2\dot\us=0\,.
\end{equation}

Рівняння~(\ref{matsyuk:9}) розглядалося в працях \cite{matsyuk:Riewe} і \cite{matsyuk:Costantelos} як
рівняння, яке описує тремтіння буцім-класичної частки.

\section{КАНОНІЧНИЙ ФОРМАЛІЗМ ДЛЯ РЕЛЯТИВІСЬКОЇ ЧА\-С\-Т\-КИ В МЕХАНІЦІ ОСТРОГРАДСЬКОГО ДРУГОГО ПО\-РЯ\-Д\-КУ.}
\subsection{Однорідний гамільтонів формалізм.}
Тут розвиваємо механізм Ґрасера--Рунда--Вайсенгофа~\cite{matsyuk:Grasser, matsyuk:Rund, matsyuk:Weyssenhoff1951} однорідного гамільтонівського подання механіки Остроградського з похідними другого порядку у виразі функції Ляґранжа для релятивіського параметрично-інваріянтного варіяційного завдання. Цей механізм найкраще надається для потреб релятивіської механіки, а також, взагалі кажучи, є особливо зручним у всіх тих випадках, коли самими рамками моделі передбачається інваріянтність щодо деякої групи перетворень, яка перемішує рівноправним чином залежні змінні з незалежними, як це й відбувається під вимогою лоренц-інваріянтности.

Нехай
\begin{equation}\label{matsyuk:pr}
pr\!:T^{r}M\setminus\{0\} \to C^{r}(1,M)
\end{equation}
означає фактор-проєкцію з многовиду ненульових швидкостей Ересмана на многовид елементів торкання $r$-го порядку під дією групи (місцевих) перетворень незалежної змінної
$\mathrm{Gl^{r}}(1,\mathbb R)$ on $T^{r}M$.
Щоразу, як тільки функція Ляґранжа $\mathcal L\!: T^{r}M \mapsto \mathbb R$
задовольняє так звані умови Цермело, вона визначає деяке параметрично-інваріянтне варіяційне завдання на $T^{r}M$. Усяке таке завдання успішно переживає згадану вище факторизацію і визначає певен пучок рівнозначних між собою присаджених (до основи $C^{0}(1,M)$=M$) 1$-форм
(або ж ляґранжевих густин), означених на волокнистому многовиді $C^{r}(1,M)$ над основою $M$.
Загальна конструкція цього механізму детально розглядалася в праці~\cite{matsyuk:DGA8},
застосування теорії пучків обґрунтована Дедекером в праці~\cite{matsyuk:Dedecker}.
Тут обмежимося випадком варіяційного завдання порядку $2$
($r=2$) і, більш того, працюватимемо в місцевих координатах щоб якомога ближче підійти до конкретної фізичної моделі. Нагадаємо позначення координат у вище згаданих многовидах: $x^{\alpha}$, $u^{\alpha}$, $\dot
u^{\alpha}$, $\ddot u^{\alpha}$, $\dddot u^{\alpha}$ для $T^{4}M$
та $x^{\so}$, $x^{i}$, $\sv^{i}$, $\sv'^{i}$, $\sv''^{i}$,
$\sv'''^{i}$ для $C^{4}(1,M)$.
Як тільки у наших фізичних застосуваннях многовид $M$ перетворюється в простір-час спеціяльної теорії відносности з діягональною метрикою $(1, -1, -1, -1)$, ми впроваджуємо векторні позначки за взірцем $u=(u_{\so}, \bu)$, $u\cdot u=u_{\so}^{2}+\bu\2$,
$\bu\2=\bu\bcd\bu=u_{\alpha}u^{\alpha}$.

Нехай $\mathcal L (x, u, \dot u)$ є функцією Ляґранжа, означеною на многовиді $T^{2}M$, яка задовільняє умови Цермело:
\begin{equation}\label{matsyuk:Zermelo}
\begin{gathered}
u^\alpha \dfrac{\partial \mathcal L}{\partial \dot u^\alpha} \equiv 0 \\
u^\alpha \dfrac{\partial \mathcal L}{\partial u^\alpha} +
2\,\dot u^\alpha \dfrac{\partial \mathcal L}{\partial \dot u^\alpha} -\mathcal L\equiv
0\,.
\end{gathered}
\end{equation}
Як ведеться, застосуємо перетворення Лежандра
\[Le\!:(x, u, \dot u, \ddot u) \mapsto (x, u, \wp, \Wpp)\]
\begin{equation}\label{matsyuk:cal p}
\begin{gathered}
\Wpp=\dfrac{\partial \mathcal L}{\partial \dot u} \\
\wp=\dfrac{\partial \mathcal L}{\partial u} - \mathcal D_\tau \Wpp\,,
\end{gathered}
\end{equation}
де
\begin{equation}\label{matsyuk:Dtau}
\mathcal D_\tau = u \frac{\partial }{\mathstrut\partial x} +
\dot u \frac{\partial }{\mathstrut\partial u} +
\ddot u \frac{\partial }{\mathstrut\partial \dot u}
\end{equation}
означатиме оператор повної похідної. Зауважимо також, що в подальших застосуваннях відсутньою буде будь-яка залежність від просторово-часової змінної $x$, оскільки нас зобов'язуватиме лже-евклідівська симетрія.

Можна бачити, що умови Цермело, коли виконуються, є рівнозначні до таких:
\begin{subequations}\label{matsyuk:Z}
\renewcommand{\theequation}{\theparentequation .\alph{equation}}
\begin{gather}
u^\alpha\Wpp_\alpha \equiv 0 \label{matsyuk:Z1}\\
u^\alpha\wp_\alpha + \dot u^\alpha\Wpp_\alpha \equiv \mathcal L\,.
\label{matsyuk:Z2}
\end{gather}
\end{subequations}

У згоді з Рундом~\cite{matsyuk:Rund}, припустимо, що існує деяка $C^2$ функція $\mathcal
H$ від чотирьох змінних $(x, u, \wp, \Wpp)$, яка не є тривіяльно постійною уздовж кожної з двох останніх змінних і яка, разом з цим, є постійною уздовж перетворення Лежандра, і ми вибираємо це постійне значення рівним~$1$ без якоїсь суттєвої втрати загальности:
\begin{equation}\label{matsyuk:H=1}
\mathcal H\circ Le\equiv 1\,.
\end{equation}

Як показано у праці~\cite{matsyuk:Rund} (гляди також~\cite{matsyuk:Grasser}),
за умови
\begin{equation}\label{matsyuk:rank}
\text{rank}
\left\|\frac{\partial^2 \mathcal L}{\partial \dot u^\alpha \partial \dot
u^\beta}\right\| = \dim M-1\,,
\end{equation}
мають існувати невизначені множники $\lambda$
та $\mu$, взагалі кажучи, залежні од $x, u, \dot u, \ddot u$,
такі, що наступна {\it канонічна система} диференційних рівнянь першого порядку щодо змінних  $x, u, \wp, \Wpp$ задовольняється уздовж кожної з екстремалей варіяційного завдання з функцією Ляґранжа $\mathcal L$:
\begin{subequations}\label{matsyuk:H}
\renewcommand{\theequation}{\theparentequation .\roman{equation}}
\begin{align}
\frac{dx}{d\tau}&=\lambda\,\frac{\partial
\mathcal H}{\partial \wp} \label{matsyuk:Hx}\\
\frac{du}{d\tau}&=\lambda\,\frac{\partial
\mathcal H}{\partial \Wpp} + \mu u \label{matsyuk:Hu}\\
\frac{d\wp}{d\tau}&=-\lambda\,\frac{\partial
\mathcal H}{\partial x} \label{matsyuk:Hp}\\
\frac{d\Wpp}{d\tau}&=-\lambda\,\frac{\partial
\mathcal H }{\partial u} - \mu \Wpp \,. \label{matsyuk:Hp'}
\end{align}
\end{subequations}

Тепер розвиток довільної функції $f$ від змінних фазового простору $x$, $u$, $\wp$, $\Wpp$ задається дужкою Пуасона
\[
\big\{f,\mathcal H\big\}\overset{\mathrm{def}}= \dfrac{\partial f}{\partial x^\alpha}
        \dfrac{\partial \mathcal H}{\partial \wp_\alpha} +
 \dfrac{\partial f}{\partial u^\alpha}
        \dfrac{\partial \mathcal H}{\partial \Wpp_\alpha} -
 \dfrac{\partial f}{\partial \wp_\alpha}
        \dfrac{\partial \mathcal H}{\partial x^\alpha} -
 \dfrac{\partial f}{\partial \Wpp_\alpha}
        \dfrac{\partial \mathcal H}{\partial u^\alpha}
\]
ось яким чином~\cite{matsyuk:Grasser}
\begin{equation}\label{matsyuk:Poisson}
\dfrac{df}{d\tau} = \lambda \big\{f,\mathcal H \big\}
+ \mu \left[u^\alpha\dfrac{\partial f}{\partial u^\alpha} -
\Wpp_\alpha \dfrac{\partial f}{\partial \Wpp_\alpha}\right]\,.
\end{equation}

\subsection{Отримання функції $\mathcal H$.}
Обсяг множини можливих функцій {$\mathcal H$}, які задовольняли б (\ref{matsyuk:H=1}) є немалим. Але, оскільки кожне параметрично-інваріянтне варіяційне завдання, поставлене на просторі $T^rM$, породжує відповідне йому формулювання на просторі $C^r(1,M)$, і навпаки, можна з успіхом пробувати в цій ролі відтягнене до простору $T^rM$ гамільтонівське формулювання, перед тим збудоване на просторі $C^r(1,M)$.

Поставимо варіяційне завдання на просторі $\mathbb R\times T^rM$ у вигляді
присадженої (щодо $\mathbb R$) диференційної $1$-form $\mathcal L \,
d\tau$, де $\mathcal L$ означена лише на $T^rM$ і задовільняє умови Цермело. Нехай теж $L\, dx^{\so}$ буде тим представником відповідного пучка рівнозначних присаджених (щодо $M$) диференційних $1$-форм на волокнистому многовиді $C^r(1,M)$, який, у запроваджених вище координатах, задається співвідношеннями
\begin{equation*}
\mathcal L\, d\tau  -  (L\cc pr)\, dx^{\so} = {} - (L\circ pr)\,\vartheta \,,
\end{equation*}
де
\begin{equation}\label{matsyuk:theta}
\vartheta = dx^{\so} - u^{\so}d\tau
\end{equation}
є однією з форм торкання на многовиді $J^1(\mathbb R,M)\approx \mathbb R\times TM$.
Звідсіля\nopagebreak
\begin{equation}\label{matsyuk:cal L}
\mathcal L=u^{\so}\,L\circ pr.
\end{equation}
Канонічні кількості руху запроваджуються, як звичайно:
\begin{equation}\label{matsyuk:bp}
\left\{\begin{gathered}
\Bpp=\dfrac{\partial L}{\partial \bv'} \\
\bp=\dfrac{\partial L}{\partial \bv} - D_t \Bpp\,,
\end{gathered}\right.
\end{equation}
де
\begin{equation}\label{matsyuk:Dt}
D_t = \sv^{i} \frac{\partial }{\mathstrut\partial x^i} +
\sv'^i \frac{\partial }{\mathstrut\partial \sv^i} +
\sv''^i \frac{\partial }{\mathstrut\partial \sv'^i}
\end{equation}
означає оператор повної похідної щодо змінної $x^{\so}$.

Співвідношення поміж операторами (\ref{matsyuk:Dtau}) та (\ref{matsyuk:Dt}) повної похідної на відповідних просторах струменів, $J^2(\mathbb R, M)$ та (містечково)
$J^2(\mathbb R, \mathbb R^{\,\mathrm {dim}M-1})$
видається очевидним, як насправді воно і є: якщо $\mathsf f$ є місцевою функцією на просторі
$C^2(1,M)$, тоді
\begin{equation}\label{matsyuk:Dtau=Dt}
\mathcal D_\tau(\mathsf f\circ pr)=u^{\so}\,D_t \mathsf f\circ pr\,.
\end{equation}
Можна також отримати (\ref{matsyuk:Dtau=Dt}) безпосереднім упохідненням проєкції (\ref{matsyuk:pr}), яка, у третьому порядку, так виглядає в наших координатах:
\begin{equation}\label{matsyuk:u=v}
\left\{
\begin{gathered}
\bv\circ pr=\dfrac\bu {u_{\so}} \\
\bv'\circ pr=\dfrac{\buu}{u_{\so}^2} - \dfrac{\dot
u_{\so}}{u_{\so}^3}\,\bu\\
\bv''\circ pr=\dfrac{\buuu}{u_{\so}^3} -
3\,\dfrac{\dot u_{\so}}{u_{\so}^4}\,\buu + 3\left(\dfrac{\dot
u_{\so}^2}{u_{\so}^5} - \dfrac{\ddot u_{\so}}{u_{\so}^4}\right)\bu\,.
\end{gathered}
\right.
\end{equation}

Маючи у своєму розпорядженні співвідношення~(\ref{matsyuk:Dtau=Dt}), можемо також встановити  і співвідношення поміж парою кількостей руху $\wp=(\wp_{\so}, \bwp)$ та  $\Wpp=(\Wpp_{\so}, \bwpp)$
в~(\ref{matsyuk:cal p}), підрахованими для функції Ляґранжа
$\cal L$, даної взором (\ref{matsyuk:cal L}), з одного боку, і парою відтягнутих взад кількостей руху~(\ref{matsyuk:bp}) з иньшого боку:
{\allowdisplaybreaks
\begin{subequations}\label{matsyuk:all p=p}
\renewcommand{\theequation}{\theparentequation .\alph{equation}}
\begin{align}
\label{matsyuk:wp'0=p'0}
\Wpp_{\so}&=\uo\dfrac{\partial (L\cc pr)}{\partial \dot u_{\so}}=
        -\dfrac1{\uo^2}\,\bu\left(\dfrac{\partial L}{\partial \bv'}\cc pr\right)=
        -\dfrac1{\uo^2}\,\bu\,(\Bpp\cc pr)\,\\
\label{matsyuk:bwp'=bp'}
\bwpp&=\uo\dfrac{\partial (L\cc pr)}{\partial \buu}=
        \dfrac1{\uo}\left(\dfrac{\partial L}{\partial \bv'}\cc pr\right)=
        \dfrac1{\uo} (\Bpp\cc pr)\,
\end{align}
\begin{align}
\label{matsyuk:wp0=p0}
\begin{split}
\wp_{\so}&=L\cc pr+\uo\,\dfrac{\partial (L\cc pr)}{\partial u_{\so}}
        -\mathcal D_\tau \Wpp_\so \qquad\quad \text{правом (\ref{matsyuk:u=v}), (\ref{matsyuk:Dtau=Dt}) та (\ref{matsyuk:wp'0=p'0})}
        \\
         &=L\cc pr
         -\uo\left[
          \dfrac1{\uo^2}\,\bu\left(\frac{\partial L}{\partial \bv}\cc pr\right)
         +\dfrac{2}{\uo^3}\,\buu\left(\frac{\partial L}{\partial \bv'}\cc pr \right)
         -\dfrac{3\dot u_{\so}}{\uo^4} \,\bu\left(\dfrac{\partial L}{\partial\bv' }\cc pr\right)\right] \\
         &\phantom{=L\cc pr\;}-2\dfrac{\dot u_\so}{\uo^3}\,\bu(\Bpp\cc pr)+\dfrac{1}{\uo^2}\,\buu (\Bpp\cc pr)
         +\dfrac1{\uo}\,\bu(D_t\Bpp\circ pr)\\
         &=L\cc pr
          -\dfrac1{\uo}\,\bu\left(\frac{\partial L}{\partial \bv}\cc pr\right)
          +\dfrac{\dot u_\so}{\uo^3}\,\bu(\Bpp\cc pr)-\dfrac{1}{\uo^2}\,\buu (\Bpp\cc pr)\\
         &\mspace{360mu}+\dfrac1{\uo}\,\bu(D_t\Bpp\circ pr)\\
        &=L\cc pr -\bv\bp\circ pr -\bv'\Bpp\circ pr\,
\end{split}
\\[3\jot]
\label{matsyuk:bwp=bp}
\begin{split}
\bwp&=\uo\,\dfrac{\partial (L\cc pr)}{\partial \bu}
        -\mathcal D_\tau \bwpp \quad \text{правом (\ref{matsyuk:u=v}), (\ref{matsyuk:Dtau=Dt}) та (\ref{matsyuk:bwp'=bp'})}
        \\
         &=\uo
         \left[
          \dfrac1{\uo}\,\left(\dfrac{\partial L}{\partial \bv}\cc pr\right)
         -\dfrac{\dot u_{\so}}{\uo^3} \left(\dfrac{\partial L}{\partial\bv' }\cc pr\right)
         \right]
         +\dfrac{\dot u_\so}{\uo^2}\,(\Bpp\cc pr)-D_t\Bpp\circ pr)\\
         &=\dfrac{\partial L}{\partial \bv}\circ pr - D_t \Bpp\circ pr = \bp\circ pr\,.
\end{split}
\end{align}
\end{subequations}
}
Із (\ref{matsyuk:wp0=p0}) та (\ref{matsyuk:bwp=bp}) випливає, що
{\allowdisplaybreaks
\begin{subequations}\label{matsyuk:wp=p final}
\renewcommand{\theequation}{\theparentequation .\alph{equation}}
\begin{align}
\wp\, u&=\uo\, L\cc pr - \uo\,\bv'\Bpp\circ pr\,,\\
\intertext{тоді, як із (\ref{matsyuk:wp'0=p'0}) та
(\ref{matsyuk:bwp'=bp'}) правом (\ref{matsyuk:u=v}) випливає, що}
 \Wpp\dot u&= \uo\,\bv'\Bpp\circ pr\,,
\end{align}
\end{subequations} }
і, отже, (\ref{matsyuk:Z2}) справджується негайно.

В наступних розважаннях придержуємось теорії узагальнених гамільтонівських систем в такому поданні, як вона викладена у книзі~\cite{matsyuk:Krupkova}.
Отже ж, в наших координатах найліпше описувати розвиток системи ядром диференційної дво--форми
\begin{equation}\label{matsyuk:H form}
\omega=-dH\wedge dx^{\so}  + d\bp\wedge d\bx + d\Bpp\wedge d\bv\,,
\end{equation}
де знак зовнішнього добутку $\wedge$ містить в собі ще й згортку векторних диференційних форм за необхідності.
Хотілося б, аби й на многовиді $\mathbb R\times T^3M$ розвиток цієї самої системи задавався диференційною дво--формою подібного вигляду,
\begin{equation}\label{matsyuk:cal H form}
\varOmega=-d\,\mathcal H\wedge d\tau  + d\wp\wedge dx + d\Wpp\wedge du\,,
\end{equation}
де кількості руху $\wp$ та $\Wpp$ виводяться з функції Ляґранжа~(\ref{matsyuk:cal L}).

Як прийнято, кладемо
$H=\bp\bv+\Bpp\bv'-L$. Під цим припущенням легко підрахувати ріжницю поміж (\ref{matsyuk:cal H form}) та (\ref{matsyuk:H form}), зважаючи на співвідношення
(\ref{matsyuk:bwp'=bp'}, \ref{matsyuk:bwp=bp})
і на умови Цермело~(\ref{matsyuk:Z1}):
\begin{equation}\label{matsyuk:W-w}
\varOmega-pr^*\omega=
d(pr^*H+\wp_\so )\wedge dx^{\so}
-d\,\mathcal H\wedge d\tau\,.
\end{equation}
Хотілося б, аби ця ріжниця виявилася пропорційною до форми дотику~(\ref{matsyuk:theta}), а саме,
\begin{equation}\label{matsyuk:20mod}
\varOmega-pr^*\omega=\alpha\wedge\vartheta\,.
\end{equation}
Найпростішим способом узгодження у (\ref{matsyuk:W-w}) з
(\ref{matsyuk:20mod}) є покласти
\begin{equation}\label{matsyuk:dcalH}
d\,\mathcal H=u^{\so}d(pr^*H+\wp_{\so})
\end{equation}
та
\begin{equation}\label{matsyuk:psi}
\mathcal H=\uo pr^*H+\varPsi\,.
\end{equation}
Тепер заходимось визначати оцю функцію відхилення~$\varPsi$.
Із~(\ref{matsyuk:psi}) маємо:
\begin{equation}\label{matsyuk:pr*dH}
pr^*dH=\dfrac{d\,\mathcal H}{\uo}
+ \left(\varPsi-\mathcal H \right)\dfrac{d\uo}{\uo^2}-\dfrac{d\,\varPsi}{\uo}\,.
\end{equation}
Вистачить підставити (\ref{matsyuk:pr*dH}) у (\ref{matsyuk:dcalH}),
аби отримати співвідношення
\[
\dfrac{\mathcal H-\varPsi}{\uo}\,d\uo-\uo\,d\wp_\so ={}-d\,\varPsi\,,
\]
звідкіль стає зовсім зрозуміло, що
\[\begin{cases}
\varPsi=\uo\wp_\so+c\\\mathcal H=c\,,
\end{cases}\]
а також, що, правом (\ref{matsyuk:H=1}), $c=1$.

Отож,
\begin{equation}\label{matsyuk:cal H}
\mathcal H=\uo pr^*H+\uo\wp_\so +1
\end{equation}

\subsection{Тремтіння ({\itshape Zitterbewegung}) буцім-класичної релятивіської частки.}

Давно тому, у
1946\textsuperscript{му}~році Фріц Боп винайшов функцію Ляґранжа, що містила приспішення, для опису другого наближення за параметром запізненої дії до руху класичної частки~\cite{matsyuk:Bopp1946}.
Виглядає знаменним те, що ляґранжіяну Бопа можна надати простої та зрозумілої форми в поняттях першої кривини Френе світової ниті частки~(\ref{matsyuk:k in u})
ось яким чином:
\begin{equation}\label{matsyuk:Bopp}
\mathcal L\overset{\mathrm{def}}= a \mathcal L_r + A \mathcal L_e
        = \frac a 2 \|u\| k^2+\frac A 2 \|u\|\,,
\end{equation}
де ми приймемо, що $a\not=0$, аби узгодитися з~(\ref{matsyuk:rank}).
Оця функція Ляґранжа задовільняє умови Цермело~(\ref{matsyuk:Zermelo}).
Перший доданок у~(\ref{matsyuk:Bopp}), $\mathcal L_r$, виявляється того типу, що розглядався Рундом у~\cite{matsyuk:Rund} (теж гляди~\cite{matsyuk:Grasser}).
Другий доданок, $\mathcal L_e$, є функцією Ляґранжа вільної частки.
Згідно з (\ref{matsyuk:cal L}), відповідна місцева ляґранжева густина, означена в деякому околі на
многовиді $C^2(1,M)$, може бути виражена в координатах $x^{\so}$, $\bv$ та
$\bv'$:
\begin{multline}\label{matsyuk:L}
L\,dx^{\so}\overset{\mathrm{def}}= a L_r dx^{\so}+A L_e dx^{\so}\\
        = \frac{a}{2}\sqrt{(1+\bv\2)}
        \left(\frac{\bv'\2}{(1+\bv\2)^2}-\frac{(\bv\bcd\bv')^2}{(1+\bv\2)^3}\right)dx^{\so}
        +\frac{A}{2}\sqrt{(1+\bv\2)}dx^{\so}\,.
\end{multline}
Кількості руху (\ref{matsyuk:bp}) для цього ляґранжіяну є такими:
\[
\Bpp_r=\dfrac{\bv'}{(1+\bv\2)^{3/2}}-\dfrac{\bv\bcd\bv'}{(1+\bv\2)^{5/2}}\,\bv \\
\]
\begin{multline*}
\bp_r=-\dfrac{\bv''}{(1+\bv\2)^{3/2}}+3\,\dfrac{\bv\bcd\bv'}{(1+\bv\2)^{5/2}}\,\bv' \\
+\dfrac{\bv\bcd\bv''}{(1+\bv\2)^{5/2}}\,\bv
-\dfrac{1}{2}\dfrac{\bv'\2}{(1+\bv\2)^{5/2}}\,\bv
-\dfrac{5}{2}\dfrac{(\bv\bcd\bv')^2}{(1+\bv\2)^{7/2}}\,\bv\,.
\end{multline*}
Запроваджуємо стандартну функцію Гамільтона
\begin{multline}\label{matsyuk:serif H}
H=\bp\bv+\Bpp\bv'-L
        \\ \overset{\mathrm {def}}=
        aH_r+AH_e=a\bp_r\bv+a\Bpp_r\bv'-aL_r+A\bp_e\bv-AL_e\,,
\end{multline}
тому, що $\Bpp_e=0$. Потрібно вилучити змінну $\bv'$
із~(\ref{matsyuk:serif H}). Підраховуємо:
\[
\begin{cases}
\Bpp_r\bv'=2L_r \\
\Bpp_r\2+(\Bpp_r\bv)^2=2\,\dfrac{L_r}{(1+\bv\2)^{3/2}}\,,
\end{cases}
\]
і от остаточно маємо функцію Гамільтона
\begin{equation}\label{matsyuk:H final}
H=\bp\bv+\frac 1{2a}\left(1+\bv\2\right)^{3/2}\left(\Bpp\2+(\Bpp\bv)^2\right)
    - \frac A{2}\sqrt{1+\bv\2}\,.
\end{equation}

У праці~\cite{matsyuk:Bopp1946}, на сторінці~199, Фріц Боп стверджував: \ulqНа класичний рух накладається деяке тремтіння, яке описується новими змінними $\bv$
та $\Bpp$. Воно провадить до ефектів спінового типу \dots\urQ\nopagebreak\footnote{\ulq Der
klassischen Bewegung \"uberlagert sich eine Zitterbewegung, die
durch die neuen Variabeln $\bv$ und $\Bpp$ beschrieben wird. Sie
f\"uhrt zu spinartigen Effekten\dots\urq }

Гамільтонівську функцію на просторі $T^3M$ можна отримати з~(\ref{matsyuk:cal H}):
\begin{equation}\label{matsyuk:cal H final}
\mathcal H=\wp u+\frac {1}{2a}\|u\|^3\Wpp\2-\frac A 2 \|u\|+1\,.
\end{equation}
Зауважимо, що цей сам вираз можна було б отримати безпосередньо з припущення
\begin{equation}\label{matsyuk:cal H directly}
\mathcal H=\wp u + \Wpp\dot u - \mathcal L + 1\,,
\end{equation}
вважаючи, що $\mathcal L$ узято з~(\ref{matsyuk:Bopp}).

З огляду на~(\ref{matsyuk:rank}),
несила повністю розв'язати перетворення Лежандра~(\ref{matsyuk:cal p}). Зате
ось як можна вилучити змінну $\dot u$ з~(\ref{matsyuk:cal H directly}):
спочатку підрахуємо кількості руху для~(\ref{matsyuk:Bopp})
\begin{gather*}
 \Wpp=\dfrac{a}{\|u\|^5}\left[u\2\dot u-(u\cdot\dot u)u\right] \\
 \wp= \dfrac{Au}{2\|u\|} {}-a\left[\dfrac{\ddot u}{\|u\|^3}
 -3\,\dfrac{u\cdot\dot u}{\|u\|^5}\,\dot u
 -\dfrac{u\cdot\ddot u}{\|u\|^5}\,u+\dfrac{\dot u\2}{2\|u\|^5}\,u
 +\dfrac{5}{2}\dfrac{(u\dot u)^2}{\|u\|^7}\,u
\right].
\end{gather*}
Наступним кроком, виразимо всі величини у~(\ref{matsyuk:cal H directly}),
куди входить $\dot u$, у змінних $\Wpp$ та $u$:
\begin{equation}\label{matsyuk:}
\left\{
\begin{aligned}
\Wpp\dot u&=\dfrac{\|u\|^3}{a}\,\Wpp\2 \\
\mathcal L_r&=\dfrac{\|u\|^3}{2a^2}\,\Wpp\2\,,
\end{aligned}
\right.
\end{equation}
і, врешті, підставимо до~(\ref{matsyuk:cal H directly}), щоб остаточно отримати функцію Гамільтона~(\ref{matsyuk:cal H final}).

{\it\bf Зауваження.\ }
Наш підхід до побудови функції Гамільтона ріжниться від підходу Ґрасера.
Він швидше пов'язаний з розглядом функцій Ляґранжа, квадратичних за швидкостями, в теорії просторів Фінслера.

Тепер не важко отримати рівняння Ойлера--Пуасона четвертого порядку для варіяційного завдання з функцією Ляґранжа~(\ref{matsyuk:Bopp}), відштовхуючись від гамільтонівської системи~(\ref{matsyuk:H})
та виразу~(\ref{matsyuk:cal H final}).
Для~(\ref{matsyuk:H}) маємо:
\begin{equation*}
\left\{
\begin{aligned}
 \dfrac{dx}{d\tau}&=\lambda\, u
\\
\dfrac{du}{d\tau}&=\lambda\, \dfrac{\|u\|^3}{a}\,\Wpp+\mu\,u
\\
\dfrac{d\wp}{d\tau}&=0 \\
\dfrac{d\Wpp}{d\tau}&=\lambda\, \dfrac{A}{2}\dfrac{u}{\|u\|}
        -\lambda\, \wp-\lambda\, \dfrac{3\|u\|}{2a}\,\Wpp\2u - \mu\,\Wpp\,.
\end{aligned}
\right.
\end{equation*}
З другого рівняння отримуємо значення множника $\mu$ шляхом згортки з вектором $u$ і з наступним використанням умов Цермело~(\ref{matsyuk:Zermelo}). Маємо $\mu=\frac{u\cdot\dot u}{\|u\|^2}$.

Тільки на цій стадії маємо право накласти певні в'язі на підбір мірки уздовж світової ниті. Вибираємо натуральну мірку $s$, так, що $\us\cdot\us=1$. Отримаємо
\begin{subequations}\label{matsyuk:H Zitterbewegung u=1}
\renewcommand{\theequation}{\theparentequation.\alph{equation}}
\begin{align}
\dfrac{du}{d{s}}&=\dfrac{\Wpp}{a} \label{matsyuk:dot u u=1}\\
\dfrac{d\Wpp}{d{s}}&=\dfrac{A}{2}\,\us
        -\wp-\dfrac{3}{2a}\Wpp\2 \us \,, \label{matsyuk:.wp' u=1}
\end{align}
\end{subequations}
і тепер видно, що  $\lambda=1$ і $\mu=0$, так, що рівняння розвитку~(\ref{matsyuk:Poisson}) відновлює свій звичайний вигляд.

Далі упохіднюємо рівняння~(\ref{matsyuk:dot u u=1}) і підставляємо туди рівняння~(\ref{matsyuk:.wp' u=1}), щоб отримати
\begin{gather}
\ddot \us =   \dfrac{A}{2a}\,\us
                     -\dfrac{\wp}{a}-\dfrac{3}{2a^2}\Wpp\2 \us \,,
 \label{matsyuk:..u}
\\
\dfrac{\wp\dot \us }{a}={}- \ddot \us \cdot \dot \us \,,\label{matsyuk:..u.u}
\end{gather}
а з иньшого боку, згортка (\ref{matsyuk:dot u u=1} з
(\ref{matsyuk:.wp' u=1}) дає
\begin{equation}\label{matsyuk:.wp'wp' u=1}
\Wpp\cdot \dot\Wpp={}-a\,\wp\dot \us \,.
\end{equation}
Ще одне упохіднення рівняння (\ref{matsyuk:..u}) дає
\begin{equation}\label{matsyuk:...u}
\dddot \us = \dfrac{A}{2a}\,\dot \us {}-\dfrac{3}{a^2}\,(\Wpp\cdot \dot\Wpp)\, \us
        - \dfrac{3}{2a^2}\,\Wpp\2\dot \us \,,
\end{equation}
куди ми і підставляємо (\ref{matsyuk:.wp'wp' u=1}), (\ref{matsyuk:dot u u=1}),
і, послідовно,
(\ref{matsyuk:..u.u}), аби врешті добути остаточне рівняння руху четвертого порядку
\begin{equation}\label{matsyuk:...u final}
\dddot \us + \left(\dfrac{3}{2}\,\dot \us \2 - \dfrac{A}{2a}\right)\,\dot \us
        +3\,(\dot \us \cdot \ddot \us)\, \us =0\,.
\end{equation}
Правом~(\ref{matsyuk:k2=dotu2}) на зв'язаному підмноговиді постійної величини релятивіського приспішення $k=k_\so$ рівняння~(\ref{matsyuk:...u final}) {\em зводиться до рівняння  гвинтової світової ниті руху частки з буцім-класичним спіном~(\ref{matsyuk:9}),} якщо покласти
\begin{equation*}
\omega^2=\frac{3}{2}\,k_\so^2-\frac{A}{2a}\,.
\end{equation*}

\section{ОБГОВОРЕННЯ.}
\begin{enumerate}
\item
Рівняння~\ref{matsyuk:9} було відоме ще Рівові~\cite{matsyuk:Riewe}, та його виведення безпосередньо з~(\ref{matsyuk:Dixon}), або ж з рівнянь Матісона--Папапетру~\cite{matsyuk:18}, як видається, не було очевидним.
\item
Правом формули $kk_2k_3=\lVert\us\dot\us\ddot\us\dddot\us\rVert$, яка виказує співвідношення поміж послідовними кривинами Френе у натуральному в\'ідмірі, негайно бачимо, що усі екстремалі варіяційного завдання~(\ref{matsyuk:Bopp}) мають нульову третю кривину, що, з огляду на означення світової ниті, означає, що в просторі частка переміщається у (двовимірній) площині. Варто порівняти цей результат з подібним результатом роботи~\cite{matsyuk:Yakupov}.
\item
Як давно доведено~\cite{matsyuk:Diss}, кожна з кривин Френе, взята у ролі функції Ляґранжа, дає екстремалі, вздовж яких ця сама кривина є постійною. Цей факт також спостережений Ародзем стосовно тільки першої кривини~\cite{matsyuk:9}. Однак проблема варіяційного опису стежок, усі кривини яких одночасно зберігаються, залишається відкритою.
\item
З фізичного погляду, цікавим є той факт, що рівняння~(\ref{matsyuk:Dixon}) у їх диференційних продовженнях покривають, як рівняння Матісона--Па\-па\-пет\-ру частки зі спіном, так і рівняння Лоренца--Дірака самовипромінюючої частки у згоді з передбаченнями Барута~\cite{matsyuk:ConfGRG1986, matsyuk:26}.
\item\label{enumerate5}
Слідкуючи ідеями Скоробогатька~\cite{matsyuk:6}, я свого часу
отримав (гляди \cite[с.~18]{matsyuk:V.Ya.Skorobohat'ko}, \cite[с.~88]{matsyuk:Novikov}) деякі неточкові (інтеґральні) перетворення простору--часу, які залишають незмінним точний вираз інтеґрала дії
\begin{equation}\label{matsyuk:16}
\int\mathcal L_\epsilon=\int\sqrt{\epsilon^2ds^2-d\alpha^2}\,,
\end{equation}
де $d\alpha$ вимірює поворот дотичної до світової ниті у відповідності з приростом натуральної мірки (власного часу) $ds$ вздовж неї, так що кривина виражається формулою\ \ $k=\frac{d\alpha}{ds}$.\ \ Робилися спроби надати цим нелокальним, лінійним за $\alpha$ і $s$, перетворенням фізичного значення перетворення координат при переході поміж системами відліку, які взаємно рівноприспішуються. Трактуючи змінні $\alpha$ і $s$ чисто формально, як незалежні величини, можна показати, що варіяції функції дії~(\ref{matsyuk:16}) приведуть до екстремалей постійної кривини (тобто, світових нитей рівноприспішених часток). З иньшого боку, більш детальне вивчення природи функції Ляґранжа
\begin{equation}\label{matsyuk:17}
\mathcal L_\epsilon=\sqrt{\epsilon^2-k^2}
\end{equation}
негайно провадить до концепції {\it максимального приспішення}~\cite{matsyuk:Novikov,matsyuk:Scarpetta}. Окремі автори надають різних значень цій фізичній константі. Зокрема, вкажемо на такі два значення,
що не відрізняються порядком: $\epsilon=c^{7/2}G^{-1/2}\hbar^{1/2}=6\cdot10^{53}\,\mathrm{cm/sec^2}$
(Комарницький~\cite{matsyuk:28}) та $\epsilon=5\cdot10^{53}\,\mathrm{cm/sec^2}$ (Скарпета~\cite{matsyuk:Scarpetta}).
\item
Виникають дві перепони для повного узгодження викладених в пункті~\ref{enumerate5} міркувань:
\begin{itemize}
\item[-]
по-перше, функція Ляґранжа~\ref{matsyuk:17}, потрактована, як ляґранжіян, що насправді містить вищі похідні (приспішення), вже у двовимірному випадку дає такі варіяційні рівняння Ойлера--Пуасона, серед розв'язків яких лиш пр\'ості лінії мають постійну кривину;
\item[-]
по-друге, варіяційне завдання~(\ref{matsyuk:16}) не є параметрично-ін\-ва\-рі\-я\-н\-т\-ним, оскільки функція Ляґранжа~$\lVert u\rVert\mathcal L_\epsilon$ з кривиною~$k$, яка задається виразом~(\ref{matsyuk:k in u}), не задовільняє умови Цермело~(\ref{matsyuk:Zermelo}).

\end{itemize}
{\it Ляґранжіян~(\ref{matsyuk:Bopp}) вільний од цих недоліків.}
\end{enumerate}



\bigskip

\begin{center}
{\bf RELATIVISTIC TOP IN THE OSTROHRADS'KYJ DYNAMICS}

\bigskip

{\it Roman~MATSYUK}

\medskip
Institute for Applied Problems in Mechanics and Mathematics\\3$^{\mbox b}$~Naukova~St., L\kern-.25em'viv, Ukraine

\end{center}

\medskip

A variational equation of the fourth order for the free relativistic top is developed starting from the Dixon's system of  equations for the motion of the relativistic dipole. The obtained equation is then cast into the homogeneous space--time Hamiltonian form.

\end{document}